\documentclass[review,10pt,number,1p]{elsarticle}
\usepackage[utf8]{inputenc}
\usepackage{float}
\usepackage{graphics}

\usepackage[footnotesize]{caption2} 

\usepackage{threeparttable} 
\usepackage{multirow}
\usepackage{flushend} 
\usepackage{verbatim}
\usepackage{graphicx}
\usepackage{color}
\usepackage{amsmath,amssymb}
\usepackage{amsmath}
\usepackage{hyperref}
\usepackage[left]{lineno}
\biboptions{sort&compress}
\DeclareUnicodeCharacter{00A0}{~}

\begin{document}
\title{Increasing the rate capability for the cryogenic stopping cell of the FRS Ion Catcher}

\author[a,b]{J.W. Zhao\corref{cor}}
\cortext[cor]{Corresponding author.} 
\author[c]{D. Amanbayev}
\author[b,c]{T. Dickel}
\author[c]{I. Miskun}
\author[b,c]{W. R. Plaß}
\author[b,d]{N. Tortorelli}

\author[b,c]{S. Ayet San Andr{\'e}s}
\author[b,c]{S{\" o}nke Beck}
\author[c]{J. Bergmann}
\author[ge]{Z. Brencic}
\author[e]{P. Constantin}
\author[b]{H. Geissel}
\author[c]{F. Greiner}
\author[c]{L. Gröf}
\author[b]{C. Hornung}
\author[b]{N. Kuzminzuk}
\author[c]{G. Kripk{\'o}-Koncz}
\author[f,h]{I. Mardor}
\author[b,i]{I. Pohjalainen}
\author[b,c]{C. Scheidenberger}
\author[d]{P. G. Thirolf}

\author[l]{S. Bagchi}
\author[b]{E. Haettner}
\author[b]{E. Kazantseva}
\author[b]{D. Kostyleva}
\author[e]{A. Oberstedt}
\author[b]{S. Pietri}
\author[j]{M. P. Reiter}
\author[b]{Y. K. Tanaka}
\author[k]{M. Wada}

\author[e]{D. L. Balabanski}
\author[f]{D. Benyamin}
\author[m]{M. N. Harakeh}
\author[b]{N. Hubbard}
\author[m]{N. Kalantar-Nayestanaki}
\author[b,c]{A. Mollaebrahimi}
\author[b]{I. Mukha}
\author[b,m]{M. Narang}
\author[k]{T. Niwase}
\author[n]{Z. Patyk}
\author[b]{S. Purushothaman}
\author[e]{A. Rotaru}
\author[e]{A. Sp{\v a}taru}
\author[q]{G. Stanic}
\author[ge]{M. Vencelj}
\author[b]{H. Weick}
\author[b]{J. Yu}
\author{the Super-FRS Experiment Collaboration}

\address[a]{State Key Laboratory of Nuclear Physics and Technology, School of Physics, Peking University, Beijing, 100871, China}
\address[b]{GSI Helmholtzzentrum für Schwerionenforschung, 64291 Darmstadt, Germany}
\address[c]{II. Physikalisches Institut, Justus-Liebig-Universität Gießen, 35392 Gießen, Germany}
\address[d]{Ludwig-Maximilians-Universität München, Am Coulombwall 1, 85748 Garching, Munich, Germany}
\address[ge]{Jozef Stefan Institute, SI-1000, Ljubljana, Slovenia}
\address[e]{Extreme Light Infrastructure - Nuclear Physics, Horia Hulubei National Institute for Physics and Nuclear Engineering,  Bucharest-Magurele, 077125, Romania}
\address[f]{Tel Aviv University, 6997801 Tel Aviv, Israel}

\address[h]{Soreq Nuclear Research Center, 8180000 Yavne, Israel}
\address[i]{University of Jyv{\"a}skyl{\"a}, FI-40014, Jyv{\"a}skyl{\"a}, Finland} 
\address[l]{Department of Physics, Indian Institute of Technology (Indian School of Mines) Dhanbad, Jharkhand - 826004, India}
\address[j]{University of Edinburgh, EH8 9AB Edinburgh, United Kingdom}
\address[k]{Wako Nuclear Science Center, Institute of Particle and Nuclear Studies, High Energy Accelerator Research Organization (KEK), Wako, Saitama 351-0198, Japan 
}

\address[m]{ESRIG, University of Groningen, 9747 AA Groningen, The Netherlands}
\address[n]{National Centre for Nuclear Research, ul. Pasteura 7, PL-02-093 Warsaw, Poland
}

\address[q]{Johannes Gutenberg-Universität Mainz, 55099 Mainz, Germany}


\begin{abstract} 
At the FRS Ion Catcher (FRS-IC), projectile and fission fragments are produced at relativistic energies, separated in-flight, energy-bunched, slowed down, and thermalized in the ultra-pure helium gas-filled cryogenic stopping cell (CSC). Thermalized nuclei are extracted from the CSC using a combination of DC and RF electric fields and gas flow. This CSC also serves as the prototype CSC for the Super-FRS, where exotic nuclei will be produced at unprecedented rates 
making it possible to go towards the extremes of the nuclear chart. Therefore, it is essential to efficiently extract thermalized exotic nuclei from the CSC under high beam rate conditions, in order to use the rare exotic nuclei which come as cocktail beams. The extraction efficiency dependence on the intensity of the impinging beam into the CSC was studied with a primary beam of $^{238}$U and its fragments. Tests were done with two different versions of the DC electrode structure inside the cryogenic chamber, the standard 1 m long and a short 0.5 m long DC electrode. In contrast to the rate capability of 10$^4$ ions/s with the long DC electrode, results show no extraction efficiency loss up to the rate of 2$\times$10$^5$ ions/s with the new short DC electrode. This order of magnitude increase of the rate capability paves the way for new experiments at the FRS-IC, including exotic nuclei studies with in-cell multi-nucleon transfer reactions. The results further validate the design concept of the CSC for the Super-FRS, which was developed to effectively manage beams of even higher intensities.
\end{abstract}

\begin{frontmatter}
\end{frontmatter}

\section{Introduction}
At the FRS Ion Catcher (FRS-IC)~\cite{plass2013,plass2019},  high-precision experiments of thermalized exotic nuclei are done at the final focus of the symmetric branch of the in-flight fragment separator (FRS)~\cite{geissel1992} at the GSI Helmholtz
Center for Heavy Ion Research, Darmstadt, Germany. The FRS-IC consists of a gas-filled Cryogenic Stopping Cell (CSC)~\cite{purushothaman2013,ranjan2015}, a Radio Frequency Quadrupole (RFQ) beamline, and a Multiple-Reflection Time-Of-Flight Mass Spectrometer (MR-TOF-MS)~\cite{plass2008,dickel2015,san2019}. The CSC also serves as the prototype of the next-generation CSC~\cite{dickel2016} of the Super-FRS~\cite{geissel2003}. Exotic nuclei produced at relativistic energies by projectile fragmentation and fission are separated in-flight by the FRS. 
These nuclei are stopped in the CSC by the ultra-pure helium gas at cryogenic temperatures and are then transported via the RFQ beamline to the MR-TOF-MS for high-precision direct mass measurements.

 As a universal method to convert fast exotic beams to low-energy and low-emittance beams for precision experiments with stored ions, e.g., mass measurements and decay and laser spectroscopy~\cite{wada2013}, gas-filled stopping cells have been widely used to slow down and thermalize exotic nuclei produced by fusion-evaporation, in-flight fragmentation and fission~\cite{wada2003,savard2003,weissman2005,neumayr2006,ranjan2011,kaleja2020}. 
One of the challenges of this approach is the operation with high beam intensities as space-charge and plasma effects can deteriorate the efficiency~\cite{huyse2002,takamine2005,morrissey2007,moore2008,facina2008,varentsov2019}. The CSC of the FRS-IC is operated at cryogenic temperatures to achieve high cleanliness of the stopping gas, with strong DC fields for fast extraction, and with an RF carpet to reach the highest buffer gas density (stopping efficiency).

An ability to handle incoming beams of high intensity without deteriorating ion extraction efficiency (i.e. high rate capability) opens possibilities for new experiments where the nuclear reaction takes place in the stopping volume of the CSC.
Examples of such experiments include in-cell multi-nucleon transfer (MNT) reaction~\cite{dickel2020} and spontaneous fission~\cite{plass2019,mardor2020,washitz2023} studies.
A new dedicated DC-cage has been developed~\cite{rotaru2022} for the high-rate experiments above to surpass the limitations of a standard DC-cage.
The rate capability of both systems has been studied experimentally, and the results are presented in this paper.

\section{Experiment}
Two experiments (i.e., Experiment I and Experiment II) have been performed with a $^{238}$U primary beam to study the rate capability of the CSC at the FRS-IC in 2016 and 2021 with the experimental setup described in detail in Refs.~\cite{plass2013,purushothaman2013,ranjan2015,miskun2019}. In past investigations~\cite{reiter2016}, very short primary beam spills on the millisecond scale have been used to simulate high-rate beams. In the experiments presented here, more realistic experimental conditions with a spill length of a few seconds were used. In both experiments, the optimum range of the ions for efficiently stopping in the CSC was tuned by varying the homogeneous degrader installed in front of the CSC. Thus the ratio of stopped to injected ions is maximized. The overall efficiency of the thermalization process, given by the product of stopping efficiency and extraction efficiency, was measured for different beam intensities. The beam intensities were measured with a plastic scintillator mounted in front of the CSC. The stopping efficiencies were determined from a measurement of the range distribution and the known areal density of the stopping cell. The overall efficiencies were determined by counting the ions extracted from the CSC with the MR-TOF-MS. The extraction efficiencies could then be calculated as the ratio of overall efficiency to stopping efficiency.

In Experiment I, the extraction efficiencies were investigated for different rates of the $^{238}$U primary beam at 300 MeV/u with a spill length of 1 s. Thermalized $^{238}$U ions were extracted from the CSC and measured with the MR-TOF-MS. As uranium is one of the most reactive elements, it tends to form molecules from reactions with contaminants contained in the helium gas. The extraction efficiencies were calculated from the total count rates of all observed forms (i.e., $^{238}$U$^{2+}$, $^{238}$UO$^{2+}$, $^{238}$UOH$^{2+}$, and $^{238}$UO$_2^{2+}$) identified via high precision mass measurements with the MR-TOF-MS. In this experiment, the long DC electrode structure was used in the CSC (i.e., long DC-cage), which has a stopping volume with a length of 105.4 cm and a diameter of 25 cm. The CSC was operated with the helium areal density of $3.16\pm0.35$~mg/cm$^2$ (corresponding to the pressure of $59\pm6$ mbar at the temperature of 94 K) and a DC push field of about 20 V/cm. Three different values of the repelling RF voltage (i.e., 94 V$_\text{pp}$, 40 V$_\text{pp}$, and 28 V$_\text{pp}$) were applied to the RF carpet. This assists in verifying that the decline in extraction efficiency is attributable to space-charge effects within the bulk of the stopping volume rather than the insufficient repelling force exerted by the RF carpet, which could impact ion motion in that region. 

\begin{figure}[htb]
	\centering
	\includegraphics[width=0.5\textwidth]{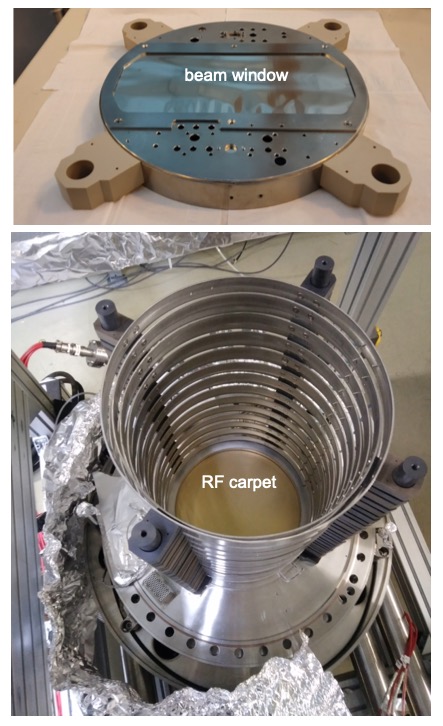}
	\caption{Photo of the short DC-cage installed on the flange of the CSC. The DC-cage consists of 27 ring electrodes with a pitch of 2 cm.  The ``golden" RF carpet and the entrance electrode (with a beam window) installed on the two ends of the DC-cage are shown as well.}
	\label{f1}
\end{figure}

A short DC electrode structure (i.e., short DC-cage shown in Fig.~\ref{f1}) with a length of 48.2 cm and a diameter of 26.7 cm was used in Experiment II. This short DC-cage shortens the extraction path and allows applying a higher DC push field in the CSC. In addition, the pitch of the short DC-cage is reduced to half of the longer one used in previous experiments and the diameter is slightly increased; together these increase the effective radius of the stopping volume by 15\% and thus the effective area by 30\%. This further increases the rate capability of the system. The extraction efficiencies were investigated with a $^{235}$U secondary beam produced by a $^{238}$U primary beam at different intensities with a spill length of 5 s and 12 s. $^{235}$U ions were produced via $^{238}$U projectile fragmentation at 1000 MeV/u in a beryllium target with an areal density of 0.664 g/cm$^2$. The target was followed by a 0.223 g/cm$^2$ Nb stripper to reach the highest charge state at this energy. $^{235}$U ions were separated in flight using twofold magnetic rigidity analysis and a 2 g/cm$^2$ Al wedge degrader located at the central focal plane of the FRS. The identification of the ions was performed using the standard particle detectors of the FRS. After further slowing down in the 
degrader at the final focal plane, the ions were injected into the CSC. 
The CSC was operated with the helium areal density of $1.12\pm0.12$~mg/cm$^2$ (corresponding to the pressure of $36\pm4$~mbar at the temperature of 75 K) and a DC push field of about 30.4~V/cm. The RF carpet was operated with 100\% transmission efficiency. In Experiment II, extraction efficiencies were calculated from the count rates of $^{235}{\rm UO}^{2+}$ measured with the MR-TOF-MS.

\section{Results and discussion}
\begin{figure}[htb]
	\centering
	\includegraphics[width=0.9\textwidth]{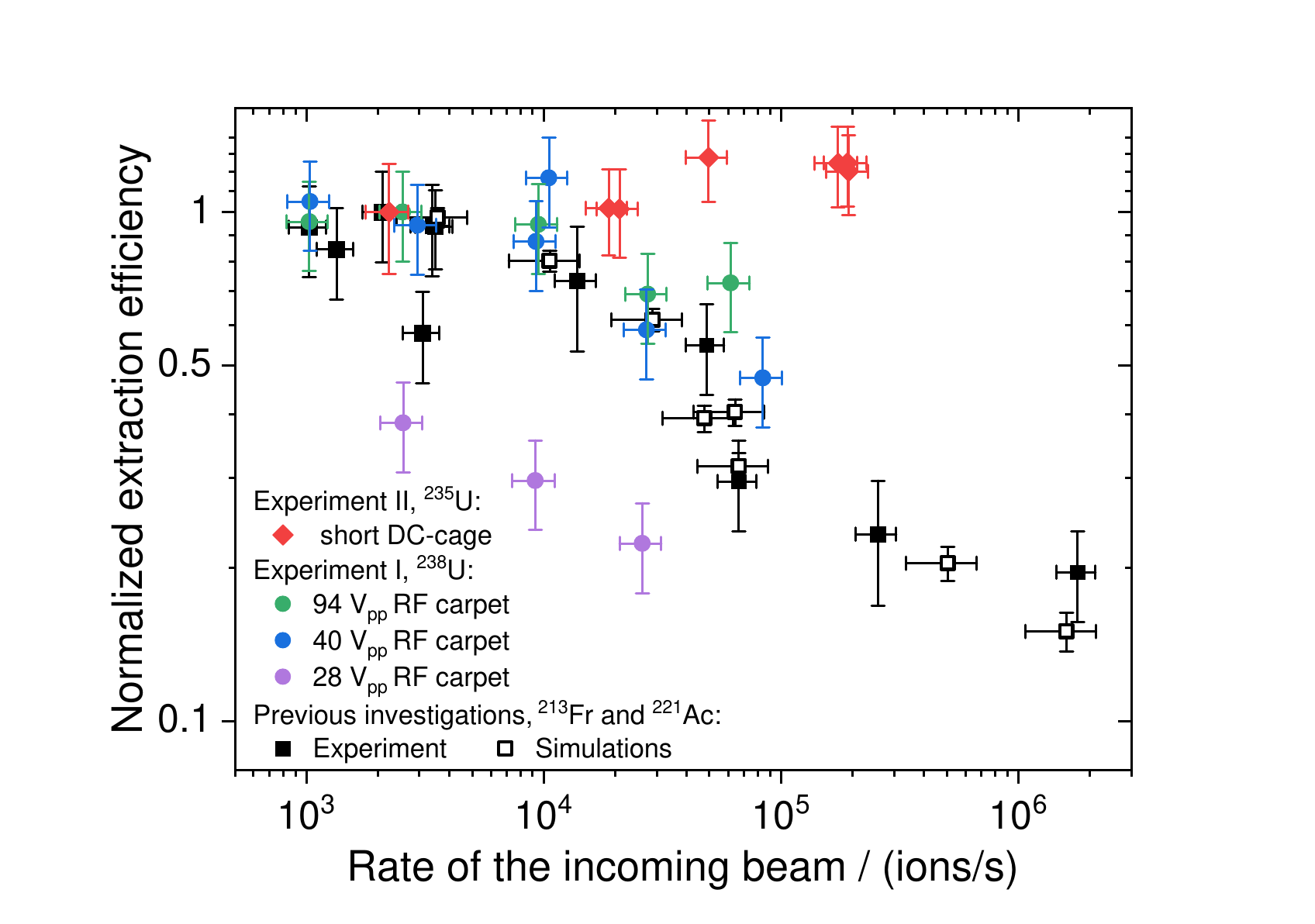}
	\caption{Rate capability achieved with the long DC-cage (filled squares and circles) and the short DC-cage (filled diamonds). Simulations (open squares) done for previous investigations with the long DC-cage agree well with experimental results (filled squares).}
	\label{f2}
\end{figure}

 In order to eliminate effects from the stopping in the CSC and transporting from CSC to the MR-TOF-MS,  
 the extraction efficiencies are normalized to a rate at which the CSC has the full extraction efficiency. For 
Experiment I, all data points were normalized to the extraction efficiency measured with the beam intensity of 2.5$\times$10$^3$ ions/s and the repelling RF voltage of 94 V$_\text{peak-peak}$ applied to the RF carpet. The normalization of data from Experiment II was done with the extraction efficiency measured at the beam intensity of 2.2$\times$10$^3$ ions/s. Figure~\ref{f2} shows a compilation of the results of past investigations done in 2014~\cite{reiter2016} and new measurements performed in the present work.

The rate capabilities were found to be independent of the repelling RF voltage applied to the RF carpet. For low rates, the resulting transmission efficiencies of the three different repelling RF voltages (i.e., 94 V$_\text{pp}$, 40 V$_\text{pp}$, and 28 V$_\text{pp}$) are 100\%, 100\%, and 40\%, respectively. As shown (by filled circles) in Fig.~\ref{f2}, the decreasing trend of the extraction efficiency is independent of the repelling RF voltage applied to the RF carpet.  As seen from the figure, the new measurements (filled blue and green circles) performed with the long DC-cage agree with the simulation results (open squares) and the former experimental data with short beam spills (filled squares). Up to the incoming beam rate of about 10$^{4}$ ions/s of $^{238}$U injected into the CSC, the extraction efficiency stays constant. However, the extraction efficiency decreases quickly at higher rates.  This behavior has been understood by space-charge effects in the previous investigations~\cite{reiter2016}. With increasing the beam rate, these effects cause a severe deflection of the thermalized ions towards the DC electrode, thus only ions stopped in the region close to the RF carpet can be efficiently extracted. 

In contrast to the long DC-cage, the extraction efficiency is significantly improved with the short DC-cage and there is no loss in extraction efficiency up to an incoming beam rate of 2$\times$10$^5$ ions/s as shown with the filled diamonds in Fig.~\ref{f2}. Simulations show the rate limit can reach 10$^7$ - 10$^8$ ions/s~\cite{rotaru2022}. However, higher beam intensities could not be tested as the radiation limit of the experimental cave was reached. The improvement is due to (i) the smaller pitch design that reduces the near-field distortions and allows efficient transporting of the ions stopped closer to the DC electrode, (ii) the larger diameter of the DC electrode that enlarges the stopping volume of the CSC in the radial direction, (iii) the higher DC push field that leads to faster removal of He ion–electron pairs and (iv) the shorter DC-cage that reduces the impact time of the space-charge on the ion extraction. 
The short DC-cage design (i.e., the smaller pitch and the larger diameter of the DC electrode) increases the stopping volume in the radial direction, which is crucial for efficient stopping and extracting the energetic fission and MNT fragments produced inside the CSC. 

The results of Experiment I validate the simulation model used to project the rate capability of the next-generation CSC for the Super-FRS~\cite{dickel2016}.
Compared to the present CSC with the standard (long) DC-cage, the next-generation CSC will exhibit notable enhancements in terms of 2 times shorter ion paths and 3 times stronger electric fields.
Additionally, multiple RF carpets will cover a larger area relative to the stopping volume.
According to the simulations, these advancements combined will further reduce ion losses caused by space-charge effects, enabling the next-generation CSC to meet the required rate capability of $10^7$~ions/s.
A configuration of the present CSC with the short DC-cage is aligned closely with the design concepts of the next-generation CSC.
Therefore, the successful use of the short DC-cage in Experiment II surpassing the rate capability of the long DC-cage by more than an order of magnitude provides additional support for the selected design concepts.

\section{Conclusions and outlook}

The rate capability of the CSC at the FRS-IC has been studied with the $^{238}$U primary beam and its projectile fragments with a spill length scale of seconds. With the standard long DC-cage, the extraction efficiency decreases for a beam rate higher than 10$^{4}$ ions/s. In contrast, no extraction efficiency loss is observed up to a rate of 2$\times$10$^{5}$ ions/s with the newly developed short DC-cage. This new rate capability is achieved by employing a shorter and wider DC electrode structure, which has better tolerance of the space-charge effects. This paves the way for exotic nuclei studies at the FRS-IC with in-cell multi-nucleon transfer reactions. The new results not only provide experimental confirmation of the advantages of the CSC with the short DC-cage, but also complete the validation of the simulation model and justify its use for the next-generation CSC~\cite{dickel2016} of the Super-FRS at FAIR. 

\section*{Acknowledgments}
We thank K.-H. Behr, M. Will, P. Schwarz, T. Weber, and B. Szczepanczyk for excellent technical support. The results presented here are based on the experiment S530, which was performed at the FRS at GSI in the context of FAIR Phase-0. Financial support was provided by the GSI/LMU R\&D project LMTHI2023, 
by the Hessian Ministry for Science and Art (HMWK) through the
LOEWE Center HICforFAIR, by HGS-HIRe, by the Justus-Liebig-Universität Gießen and GSI under the JLU-GSI strategic Helmholtz partnership agreement, and by contract PN 23.21.01.06 sponsored by the Romanian Ministry of Research, Innovation and Digitalization. The paper was partly financed by the international project “PMW” of the Polish Minister of Science and Higher Education, active in the period of 2022-2024, grant Nr 5237/GSI-FAIR/2022/0. N. K.-N. acknowledges the support, for his stay at GSI, by the ExtreMe Matter Institute EMMI at the GSI Helmholtzzentrum fuer Schwerionenphysik, Darmstadt, Germany.

\nolinenumbers


\bibliographystyle{elsarticle-num}

\end{document}